\documentclass[%
 reprint,
 amsmath,amssymb,
 aps,
 prb,
]{revtex4-2}
\usepackage{xcolor}

\usepackage{graphicx}
\usepackage{dcolumn}
\usepackage{bm}
\usepackage{bbold}
\usepackage{appendix}
\usepackage{hyperref}

\usepackage[normalem]{ulem}

\usepackage{caption}
\usepackage{subcaption}

\usepackage{cancel}

\newcommand{\toadd}[1]{{#1}}
\newcommand{\toremove}[1]{\ignorespaces}

\begin{document}

\title{A framework for efficient ab initio electronic structure with Gaussian Process States}

\author{Yannic Rath}
\email{yannic.rath@kcl.ac.uk}
\affiliation{Department of Physics, King’s College London, Strand, London WC2R 2LS, United Kingdom}%

\author{George H. Booth}
\email{george.booth@kcl.ac.uk}
\affiliation{Department of Physics, King’s College London, Strand, London WC2R 2LS, United Kingdom}%

\date{\today}

\begin{abstract}
    We present a general framework for the efficient simulation of realistic fermionic systems with modern machine learning inspired representations of quantum many-body states, towards a universal tool for {\em ab initio} electronic structure.
    These machine learning inspired ansatzes have recently come to the fore in both a (first quantized) continuum and discrete Fock space representations, where, however, the inherent scaling of the latter approach for realistic interactions has so far limited practical applications. 
    With application to the `Gaussian Process State', a recently introduced ansatz inspired by systematically improvable kernel models in machine learning, we discuss different choices to define the representation of the computational Fock space. We show how local representations are particularly suited for stochastic sampling of expectation values, while also indicating a route to overcome the discrepancy in the scaling compared to continuum formulated models. 
    We are able to show competitive accuracy for systems with up to 64 electrons, including a simplified (yet fully {\em ab initio}) model of the Mott transition in three-dimensional hydrogen, indicating a significant improvement over similar approaches, even for moderate numbers of configurational samples.
\end{abstract}
\maketitle
\makeatletter

\section{Introduction}
Interactions of many electrons with each other and their environmental nuclear potentials give rise to almost all the complexity in chemical and materials science.
Accurate simulations of these quantum particles with their known interactions can describe emergent properties of a system, and are therefore a key challenge linking the electronic scale and trustworthy predictions of relevant physical observables from fundamental physical principles.
However, these quantum problems are formulated in a Hilbert space which inherently scales exponentially with number of particles, and hence numerically tractable many-body wave function approaches typically require approximations, effectively compressing the information in this space.

Many of these approximate wave function approaches are based around explicitly imposing an appropriate functional form of the many-electron state.
These well-established representations are generally directly informed by exploiting some physical characteristics or intuition of the state which is exploited in order to describe them compactly, such as Laughlin~\cite{Laughlin1983}, BCS~\cite{Bardeen1957} or Gutzwiller~\cite{Gutzwiller1963} states.
However, since the structure of the target state depends specifically on the underlying physics of the studied system, most introduced state approximations for many-electron wavefunctions are not universally suitable for all systems of interest.
This makes these representations successful for specific classes of systems.
This approach can also encompass more flexible forms which are still nevertheless restricted in their applicability, such as coupled cluster wavefunctions~\cite{RevModPhys.79.291} (which require low-rank correlations) or tensor networks~\cite{ORUS2014117} (which require low entanglement).

The framework of variational Monte Carlo (VMC)~\cite{becca_sorella_2017} makes it possible to use any functional form as a model for the many-body wave function, as long as it can be efficiently sampled in a chosen computational basis.
Optimization of its parameters and extraction of many-body expectation values of interest are then enabled through efficient stochastic sampling of the configuration space.
In recent years, this has enabled wave function models inspired by classical machine learning (ML) to come to the fore due to their ability to describe complicated functions of many variables in a black-box and efficient fashion~\toremove{\cite{Carleo2018, PhysRevX.7.021021, Gao2017, Nomura2017, Huang2021, Glasser2018, Jia2019}}\toadd{\cite{doi:10.1126/science.aag2302,Carleo2018, PhysRevX.7.021021, Gao2017, Nomura2017, Huang2021, Glasser2018, Jia2019}}.
Importantly, such ML models, e.g., neural network architectures or kernel models, are typically not limited by rigid imposed functional forms, and in principle can be improved systematically to arbitrary accuracy to model the many-body correlations in the quantum state.
Although the speed of this convergence in desired expectation values with the complexity of the model is not guaranteed, the systematic and unbiased ability to describe many-body effects without restriction in rank or range represents probably the most important advantage over other established models typically used in VMC such as (Slater-)Jastrow ansatzes~\cite{Jastrow1955, becca_sorella_2017}.

With increasingly many successful applications of ML inspired models for quantum many-body wave functions often challenging the state-of-the-art~\cite{PhysRevX.11.031034, Roth2022}, this route is considered a promising candidate for a truly universal quantum many-body method.
In this work, we build on the Gaussian Process State (GPS)~\cite{Glielmo_2020, Rath_2020, Rath_2022}, an ML-inspired wave function model motivated from Bayesian kernel models.
This ansatz takes a particularly simple form which has been shown to practically reach accuracies comparable to similarly systematically improvable neural network inspired states.
We apply the state to electronic problems interacting with the physical long-ranged Coulomb interaction as a step towards realistic electronic structure, and discuss the challenges which arise in this context for VMC methods in a Fock space picture.
We ameliorate many of these difficulties when sampling dense Hamiltonians in this representation by combining the GPS with approaches previously considered in the orbital-space VMC literature~\cite{doi:10.1063/1.4829835}, in particular constructing the Fock space from localized basis functions.
This also paves the way to achieve a reduced (asymptotic) scaling of the method by exploiting the emerging natural sparsity of the Hamiltonian in this representation~\cite{doi:10.1063/5.0025055, Wei_2018, sabzevariFasterLowerScaling2018, Hachmann_2006}, which would bring the scaling into agreement with continuum real-space VMC~\cite{Wei_2018, foulkesQuantumMonteCarlo2001}, as has recently also been the focus of ML-inspired quantum states~\toremove{\cite{PhysRevResearch.2.033429,Hermann_2020,spencerBetterFasterFermionic2020,https://doi.org/10.48550/arxiv.2205.09438,hermannAbinitioQuantumChemistry2022, liInitioCalculationReal2022, vonglehnSelfAttentionAnsatzAbinitio2022, Han2018}}\toadd{\cite{PhysRevResearch.2.033429,Hermann_2020,spencerBetterFasterFermionic2020,https://doi.org/10.48550/arxiv.2205.09438,hermannAbinitioQuantumChemistry2022, liInitioCalculationReal2022, vonglehnSelfAttentionAnsatzAbinitio2022, Han2018, https://doi.org/10.48550/arxiv.2302.04168, https://doi.org/10.48550/arxiv.2110.05064, https://doi.org/10.48550/arxiv.2205.14962}}. This opens the possibility to practically treat larger systems.

We demonstrate the applicability of the methodology in combination with our novel state parameterization, with application to different test systems, reaching competitive energies compared to established methodologies in the investigation of electronic systems with the real Coulomb interaction.
Our results include the description of a correlation-driven metal-insulator transition in a minimal 64 atom model of a correlated hydrogen material, representing a system size beyond what has been discussed with related approaches and towards realistic materials science applications.
Furthermore, we expect the discussed developments to be able to be combined seamlessly with other ML-inspired models, including recent parametrizations specifically tailored for electronic systems~\cite{Rios, Luo2019, morenoFermionicWaveFunctions2021}.
The following section introduces the GPS ansatz to model the electronic wavefunction which we utilize in the VMC framework for \emph{ab initio} quantum chemical calculations which we outline in section~\ref{sec:ab_initio_VMC}.
Finally, we present benchmarking results for one- and three-dimensional hydrogen materials in section~\ref{sec:results}.


\section{Gaussian Process States}
\label{sec:GPS}

The general procedure for approximating the ground state of a system with VMC is conceptually simple: Having defined a functional form for the wave function in the computational basis, defining a mapping from basis states $|x \rangle$ to the configurational wave function amplitudes $\langle \Psi | x \rangle$, expectation values are evaluated by stochastic sampling from the computational basis.
Through a numerical minimization, the chosen parametrization of the state can be optimized to find a suitable approximation of the (generally unknown) many-body target state, here considered to be the electronic ground state of the \emph{ab initio} chemical system.
Key to the success is the choice of the trial wave function ansatz and its ability to faithfully represent the physics of the target state as well as exhibiting as compact a form as possible to facilitate optimization.
These properties have recently been well served by the application of traditional machine learning models which, if carefully designed, do not require a low scaling in entanglement of the target state for efficient representation~\cite{PhysRevX.7.021021, PhysRevLett.122.065301}.

Various neural network architectures have recently been applied as a model to represent \emph{ab initio} wave functions in both a first quantized~\toremove{\cite{PhysRevResearch.2.033429,Hermann_2020,spencerBetterFasterFermionic2020,https://doi.org/10.48550/arxiv.2205.09438,hermannAbinitioQuantumChemistry2022, liInitioCalculationReal2022, vonglehnSelfAttentionAnsatzAbinitio2022, Han2018}}\toadd{\cite{PhysRevResearch.2.033429,Hermann_2020,spencerBetterFasterFermionic2020,https://doi.org/10.48550/arxiv.2205.09438,hermannAbinitioQuantumChemistry2022, liInitioCalculationReal2022, vonglehnSelfAttentionAnsatzAbinitio2022, Han2018, https://doi.org/10.48550/arxiv.2302.04168, https://doi.org/10.48550/arxiv.2110.05064, https://doi.org/10.48550/arxiv.2205.14962}} and second quantized perspective~\cite{Choo_2020,https://doi.org/10.48550/arxiv.2109.12606, zhaoScalableNeuralQuantum2022, https://doi.org/10.48550/arxiv.2301.03755, Yang2020}.
In this work, we follow the latter approach, in which we construct a computational basis from Fock states identifying the electronic occupancies of a finite number of molecular orbitals.
We use a similar ML-inspired model for the wave function that was recently introduced, dubbed the Gaussian Process State (GPS)~\cite{Glielmo_2020, Rath_2020, Rath_2022}.
The GPS representation can be derived from the application of a general kernel model, as in Gaussian process regression or kernel ridge regression.
Kernel models in machine learning recast the problem into a very high dimensional feature space, at which point the data can be described via a linear model. By allowing the effective dimension of this feature space to be systematically enlarged (and, in this work, variationally optimized), the expressiveness of the GPS can be improved systematically. In this way, the GPS represents a universal approximator of a target state, not restricted to a rigid functional form or specific correlation characteristics.

Here, we utilize the recent formulation of the GPS, which can be viewed as a model supported by a set of $M$ unentangled product states as data points explicitly driving the representation~\cite{Rath_2022}.
The number of product states, $M$, in the following referred to as the `support dimension' of the model, serves as the single hyperparameter of the model controlling its complexity (and hence both its expressibility and number of parameters).
Therefore, in keeping with the approach of other ML-inspired ansatzes, it can in principle span any state in the Hilbert space as $M$ increases.

The considered GPS model associates many-body configurations with their wave function amplitudes according to a simple form, given by
\begin{equation}
    \Psi(x) = \langle x | \Psi \rangle = \exp \left(\sum_{\alpha=1}^M \prod_{i=1}^L \epsilon_{\alpha, i, x_i} \right), \label{eq:qGPS}
\end{equation}
which is specified by $M \times L \times 4$ continuous variational parameters in the tensor $\epsilon$.
Within this parametrization, each local occupancy of the $L$ spatial orbitals, denoted by $x_i$, is used as an index into the tensor of variational parameters.
The local occupancy $x_i$ can therefore take one of four values depending on whether the orbital is unoccupied, singly occupied with a spin-up/spin-down electron, or doubly occupied with electrons of both spin types.
The model generalizes a previous incarnation of the GPS, which considered a more rigid `squared-exponential' form of the kernel based on the Hamming distance metric between `classical' configurations, thereby supporting the model with fixed integer occupancies for each degree of freedom~\cite{Glielmo_2020}.
The form of Eq.~\ref{eq:qGPS} can be considered a completely flexible parametrization of such a kernel function, allowing for a fully variational identification and weighting of the dominant correlation features, and can be efficiently evaluated for arbitrary configurations of the state.


The model can also be viewed from the perspective of a tensor network state, being analogous to an exponentiated matrix product state (MPS)~\cite{ORUS2014117} for which the matrices are constrained to be diagonal.
Equivalently, the state is a CANDECOMP/PARAFAC (CP) factorization~\cite{kiersStandardizedNotationTerminology2000, koldaTensorDecompositionsApplications2009} of the log of the wave function amplitude tensor.
The `diagonal' nature of the matrices makes the expressible amplitudes independent of the orbital ordering, and the act of exponentiation yields a combination of all possible products of terms contributing to the linear combination of product states in the exponential.
This makes it possible to capture non-trivial entanglement and results in a product over the correlation features reminiscent of Correlator Product States~\cite{Changlani2009, mezzacapoGroundstatePropertiesQuantum2009}, however, without any restriction on the ranks and ranges of the correlations which are described. We show in Ref.~\onlinecite{Rath_2022} how this state can also be represented by a neural network, with a specific architecture, exposing a duality between kernel and neural network approaches which has been previously explored in the ML community~\cite{nealBayesianLearningNeural1995,lee2018deep,g.2018gaussian}.

\section{Efficient Fock-Space VMC for ab initio Fermions}
\label{sec:ab_initio_VMC}

\subsection{Electronic VMC in second quantization}
In this work, we aim to utilize the representative power of the GPS to describe the electronic ground state of molecular systems. We first review second quantized VMC for \emph{ab initio} fermions, particularly focussing on practical approaches for our specific context.
The {\em ab initio} electronic structure Hamiltonian in the Born-Oppenheimer approximation can be expressed in a discretized basis of molecular spin-orbitals according to~\cite{szaboModernQuantumChemistry2012}
\begin{equation}
    \hat{H} = \sum_{ij}^{2 L} h^{(1)}_{ij} \, \hat{c}^\dagger_i \hat{c}_j + \frac{1}{2} \sum_{ijkl}^{2 L} h^{(2)}_{ijkl} \, \hat{c}^\dagger_i \hat{c}^\dagger_k \hat{c}_l \hat{c}_j.
    \label{eq:el_structure_ham}
\end{equation}
This Hamiltonian describes the interactions of the electrons occupying the various $2L$ degrees of freedom via the creation and annihilation operators $\hat{c}^\dagger$ and $\hat{c}$ satisfying fermionic commutation relations.
The Hamiltonian matrix elements are defined via the one-electron integrals $h^{(1)}_{ij}$ capturing the single-particle contributions from their kinetic energy and interaction with the fixed external potential.
The two-electron integrals describe instantaneous electron-electron interactions via the Coulomb interaction, defined as
\begin{equation}
    h^{(2)}_{ijkl} = \int d \mathbf{r} \int d \mathbf{r}' \, \frac{\phi^\ast_i(\mathbf{r}) \phi_j(\mathbf{r}) \phi^\ast_k(\mathbf{r}') \phi_l(\mathbf{r}')}{|\mathbf{r} - \mathbf{r'}|},
    \label{eq:two-el-integrals}
\end{equation}
which are evaluated with respect to the molecular orbital functions $\phi_i(\mathbf{r})$ defined across the real space.
The spin-orbital labels $i, j, k, l$ can be understood as compound indices indexing the two-dimensional spin degree of freedom, together with the spatial degree of freedom.
Here, we work in a restricted basis of $L$ spatial orbitals, which are the same irrespective of the spin component.
In this work, the molecular orbitals are obtained as contracted functions of the underlying linear combination of atomic orbitals as defined by various tabulated quantum chemical basis sets~\cite{https://doi.org/10.1002/wcms.1123}.
With the molecular orbitals defining the computational basis of the problem, we can use the GPS, or other general ansatzes, as the model mapping the orbital occupancies of an instantaneous electronic configuration to wave function amplitudes.

We evaluate the expectation via stochastic sampling.
In particular, the variational energy of the state is computed as
\begin{equation}
E = \frac{\langle \Psi | \hat{H} | \Psi \rangle}{\langle \Psi| \Psi \rangle} = \left \langle \frac{[\hat{H} \Psi](x)}{\Psi(x)} \right \rangle_{p(x) \sim |\Psi(x)|^2}, \label{eq:energy}
\end{equation}
where the expectation value is approximated by drawing a finite number of samples according to the unnormalized probability distribution $|\langle x | \Psi \rangle|^2$.
We generate samples with standard Markov chains utilizing the Metropolis-Hastings algorithm, in which moves are proposed based on \toadd{arbitrarily ranged} single electron hops of the configuration, also ensuring that configurations are taken from the correct particle number and spin magnetization sector.
In second quantization, the local energy terms of Eq.~\ref{eq:energy} are formally evaluated as
\begin{equation}
    \frac{[\hat{H} \Psi](x)}{\Psi(x)} = \frac{\sum_{x'} {\hat H}_{x, x'} \Psi(x')}{\Psi(x)} , \label{eq:localE}
\end{equation}
where ${\hat H}_{x, x'} = \langle x | \hat{H} | x' \rangle$ denotes a Hamiltonian matrix element.
Due to the $k$-local nature of the Hamiltonian in Eq.~\ref{eq:el_structure_ham}, which contains at most quartic dependence on the fermionic operators, each local energy evaluation thus involves $\mathcal{O}[L^4]$ terms.
Excluding vanishing terms due to the particle-conservation symmetry for a fixed number of $N$ electrons, the local energy evaluation then requires the evaluation of $\mathcal{O}[N^2 \times (2L-N)^2]$ wavefunction amplitudes~\cite{doi:10.1063/1.4829835}.

The second quantization approach allows for the incorporation of the antisymmetry directly into the many-body basis (and therefore operators expressed in the basis), as achieved via the commutation relations of the constituent operators in Eq.~\ref{eq:el_structure_ham}.
The scaling above, however, contrasts with first quantized representations of quantum states (where an explicitly anti-symmetrized ansatz for the state must be imposed).
In these, the configurations directly represent an arrangement of the $N$ electrons in real space, which avoids the variational approximation in second quantized representations associated with the restriction to the fixed subspace spanned by the basis set.
Due to the fact that the real-space first-quantized Hamiltonian acting on a single configuration only considers an analytically tractable semi-local term for the one-body operators, and a quadratically-scaling electron-electron part depending on all electron pairs, the number of terms to consider in the evaluation of Eq.~\ref{eq:localE} scales only quadratically with the number of electrons, rather than $\mathcal{O}[N^2 \times (2L-N)^2]$~\cite{foulkesQuantumMonteCarlo2001}.


\subsection{Basis choice}

In the second quantized representations, as utilized in this work, there is a further freedom concerning the choice of representation of the molecular orbitals.
The choice of the molecular orbitals is not unique as any unitary single-body rotation applied to the set of orbitals $\{\phi_i\}$ yields another valid representation.
Such a change of the basis does not alter the physical observable characteristics of the (typically inaccessible) target state, whose expectation values are independent of the change.
However, it will change the amplitudes of the wave function for each configuration in the chosen computational basis.
Consequently, it likely also affects the accuracy of the ansatz (away from the exact $M\rightarrow \infty$ limit) and the rate of convergence to this exact limit, as well as the efficiency by which the space can be sampled via a stochastic Markov process.
Furthermore, symmetries of the model are often easier to exploit in symmetry-preserving representations (where the single-particle basis transforms onto itself appropriately under action of the symmetry operations), constraining the choices available.
With the ability to incorporate structure into the basis of the second quantized formulation, making an appropriate choice is thus also a key contributing factor to the overall success of the method.
\toadd{While this choice can also be automatically tuned based on variational principles~\cite{moreno2023enhancing}, here we discuss different conceptual paradigms to construct the molecular orbital basis.}

A canonical choice for the molecular orbitals can easily be obtained by applying a self-consistent mean--field method such as Hartree-Fock (HF)~\cite{szaboModernQuantumChemistry2012} which finds an orthogonal set of molecular orbitals based on the defined atomic orbitals.
By setting up the orbitals according to a mean field calculation, the resulting basis representation automatically incorporates a large degree of physical information making this the standard basis choice for a large number of electronic structure methods.
In particular, the HF wave function is then easily obtained as a single basis configuration in which the $N$ energetically lowest orbitals are occupied.

As a consequence, target wavefunctions for relatively weakly correlated systems typically exhibit a very peaked probability structure, with dominant weights only for configurations which differ by few excitations from the HF configuration.
While the sparsity of the wavefunction is a central cornerstone of the success of post-HF methods, such as coupled cluster approaches, it can be expected to cause additional difficulties for a reliable model optimization in a VMC context.
Indeed, it was noted in Ref.~\onlinecite{Choo_2020} that a key bottleneck in the application of neural network quantum states (NQS) in second quantized bases are difficulties with the VMC optimization of the state for approximations of a peaked target distribution.
This resulted in significant numbers of configurational samples being required to achieve suitable exploration of the Hilbert space to achieve the full potential of the chosen neural network ansatz.
Though the general method overall only scales linearly in the number of samples, and it is easily parallelizable over the samples, the ability to optimize the ansatz with as few samples as possible is crucial in order to scale the method up to larger systems.

While the canonical orbitals respect a natural ordering according to single particle energies, the obtained orbital functions will typically be delocalized over the physical space.
As an alternative to the canonical construction of the orbitals, in this work, we consider orbital functions constructed to fulfil locality requirements, for which we expect some practical advantages outlined below.
Different approaches have been proposed to construct an orthogonal set of localized orbital functions, $\{\phi_i\}$, commonly either based on a direct orthogonalization of the underlying atomic orbitals~\cite{lowdinNonOrthogonalityProblem1950, reedNaturalPopulationAnalysis1985, aquilanteFastNoniterativeOrbital2006}, or a numerical optimization of a locality measure~\cite{RevModPhys.32.300, edmistonLocalizedAtomicMolecular1963, pipekFastIntrinsicLocalization1989}.
Here, we consider the Boys localization scheme, an approach of the latter category.
This constructs the orbitals via minimization of the orbital width, which can equivalently be formulated as a maximization of the quantity~\cite{doi:10.1063/1.1681683}
\begin{equation}
    \mathcal{L} = \sum_i \left| \int d \mathbf{r} \, \left( \phi_i^\ast(\mathbf{r}) \, \mathbf{r} \, \phi_i(\mathbf{r}) \right) \right|^2
\end{equation}
while ensuring orthogonality of the orbital functions by only allowing optimization via unitary rotations of orthogonalized basis functions.

\begin{figure}
    \includegraphics[width=\columnwidth]{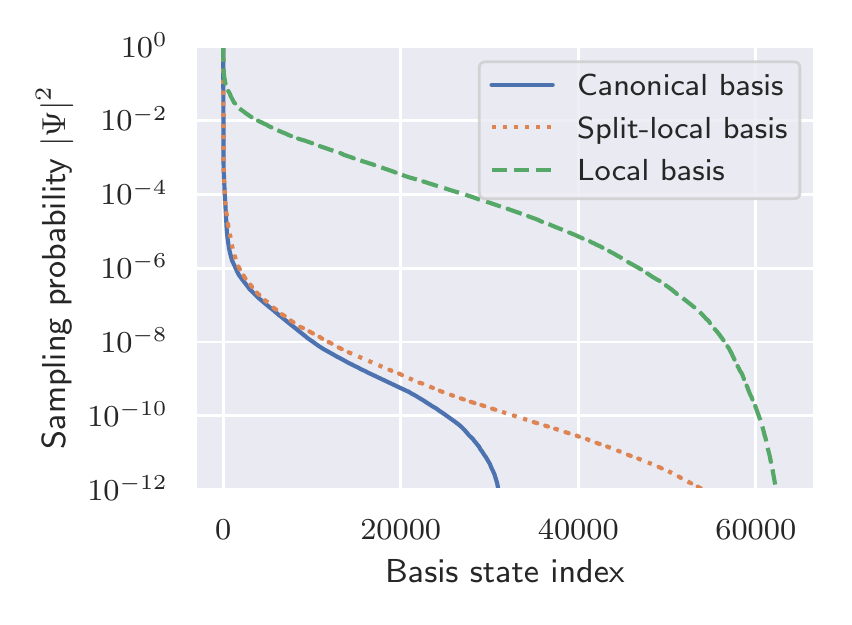}
    \caption{Ground state sample distribution of $|\Psi(\mathbf{x})|^2$, for a linear chain of ten hydrogen atoms with inter-atomic separation of $R=1.8 \, a_0$, described in a minimal basis set (STO-6G). The amplitudes associated with computational basis states are plotted in descending order from left to right with respect to different choices of the molecular orbital basis: the canonical basis from HF orbitals (blue, solid), a split-localized basis (orange, dotted), as well as a localized basis (green, dashed), according to the Boys localization criteria. Sampling probabilities are rescaled to give a value of one for the most dominant configuration, and amplitudes smaller than $10^{-12}$ are not displayed.}
    \label{fig:H10_probability_amps_distribution}
\end{figure}

In a local basis, we generally expect that the approximated wavefunction amplitudes follow a broader distribution across the computational basis, in turn improving the ability to faithfully sample expectation values required for the optimization.
This is exemplified for the ground state wavefunction of a small one-dimensional system of ten hydrogen atoms represented in Fig.~\ref{fig:H10_probability_amps_distribution}.
The figure visualizes the sampling probability distribution across the computational basis for three different basis choices.
In the canonical basis constructed from HF wavefunctions, a strongly peaked distribution becomes apparent.
For this, only $176$ basis configurations reach an occupational probability of more than $0.01 \%$ relative to that of the most strongly weighted configuration.
This strongly contrasts with the structure emerging for a localized orbital choice, giving a probability distribution spreading across a significantly broader section of the Hilbert space.
In the considered example, the localized basis gives a sampling distribution of the target state in which $27164$ basis configurations are sampled at least $0.01 \%$ as likely as the most probably sampled configuration.
Lastly, we also present the distribution for a split-localized representation in which the occupied and virtual orbitals are localized separately.
This construction is a common choice within density matrix renormalization group (DMRG) calculations due to a reduction of the orbital entanglement within regimes in which both the single-particle effects as well as the local many-body interactions contribute significantly to this entanglement~\cite{doi:10.1063/1.4905329}.
Furthermore, it still allows the mean-field state to be represented as a single configuration in the Hilbert space.
For the considered example, it still results in a sampling distribution of the target state which is concentrated around few configurations, and it is therefore not expected to help alleviate the sampling difficulties within VMC approaches.

In addition to changing the target distribution, the utilization of localized orbitals moreover paves the way for a reduction of the computational complexity of the local energy evaluation --- typically the main computational bottleneck within practical implementations.
If the molecular orbitals are sufficiently localized, many of the two-electron integrals $h^{(2)}_{i,j,k,l}$ as defined in Eq.~\eqref{eq:two-el-integrals} vanish for pairs of orbitals with large separation.
By efficiently pruning the vanishing terms from the local energy evaluation~\cite{doi:10.1063/5.0025055, Wei_2018, sabzevariFasterLowerScaling2018, Hachmann_2006}, we can obtain an asymptotic reduction to $\mathcal{O}[N^2]$ terms, thus in-line with real-space formulations of the problem.

\begin{figure}
    \includegraphics[width=\columnwidth]{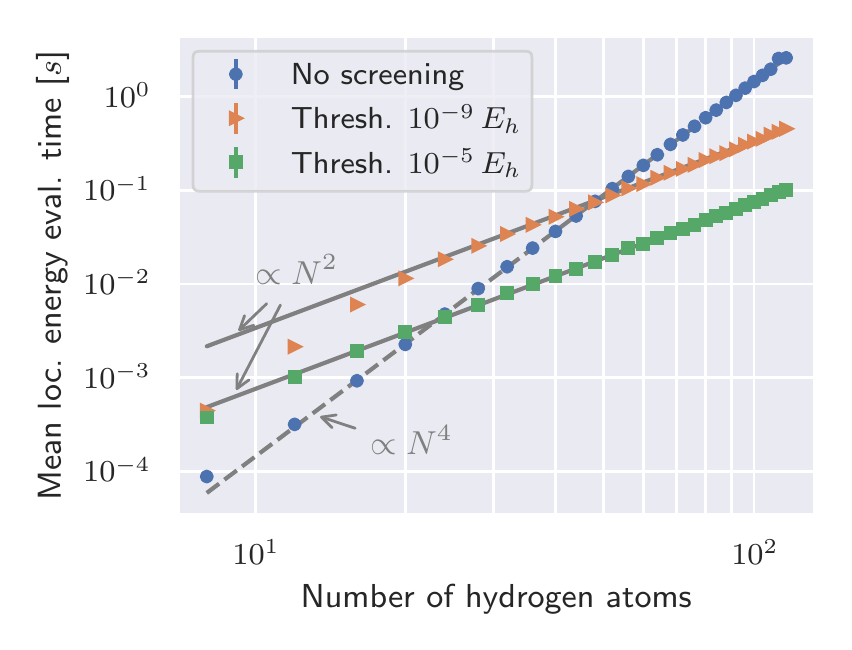}
    \caption{\toadd{Mean local energy evaluation time as a function of the number of atoms in a one-dimensional hydrogen chain (fixed inter-nuclear spacing of $1.8 \, a_0$), described in a minimal atomic basis set (STO-6G). The local energy evaluations were performed with a uniform wave function model. The figure displays the mean evaluation time with an efficient full contraction over all allowed terms (blue circles), giving an asymptotic scaling of $\mathcal{O}[N^4]$, as well as implementations with a pre-screened selection of non-vanishing contributions with pruning thresholds $10^{-5} \, E_h$ (green squares), and $10^{-9} \, E_h$ (orange triangles), resulting in an asymptotic scaling of $\mathcal{O}[N^2]$ in the chosen basis with localized orbitals. The calculations were performed on a single Intel(R) Xeon(R) Gold 6142 CPU core.}}
    \label{fig:H_chain_scaling_analysis}
\end{figure}

\toadd{To highlight the scaling reduction which can be achieved in a local basis, we report the mean evaluation time of the local energy evaluation for linear chains of hydrogen atoms of variable lengths in Fig.~\ref{fig:H_chain_scaling_analysis}.
To efficiently prune vanishing terms in the local energy evaluation, we utilize a sparse data structure listing the elements of the electron integral tensors above a chosen threshold for all possible orbital indices (one-electron terms) or index pairs (two-electron terms).
For each (single or double) electron annihilation in the application of the terms in Eq.~\eqref{eq:el_structure_ham}, the summation can then be performed by only contracting over the non-vanishing elements.
In the figure, we compare the mean evaluation time for a full contraction over all terms with the implementation exploiting the sparsity through the pre-selection of non-vanishing one- and two electron integrals.
The timings were obtained with a uniform state ansatz and do therefore not take into account a scaling dependency from the evaluation complexity of chosen wavefunction ansatzes (potentially utilizing low-rank updates for the efficient evaluation), though this will not materially affect this leading order scaling step.}

\toadd{
While the utilization of the sparse data structure comes with an additional (constant) computational overhead associated with checking the validity of an electronic move, the plot clearly confirms the asymptotic scaling of $\mathcal{O}[N^2]$ through the pre-selection of terms.
This demonstrates a clear computational advantage when the pruning is applied in a local basis as the systems increase beyond a certain size.
With an aggressive pruning threshold of $10^{-5} \, E_h$, we observe an advantage as the chains increase beyond $\approx 25$ hydrogen atoms, and for a smaller threshold of $10^{-9} \, E_h$, the crossover point is obtained at $\approx 50$ atoms.
Although the specific timing details and crossover points will be system and implementation specific, given the nuclear separation of $1.8 \, a_0$ chosen in our tests, this suggests that a computational advantage can generally be expected for the {\em ab initio} Coulomb interaction if any linear dimension of the system is of the order of $\approx 50 \, a_0$ or greater.
While the pruning of vanishing terms was implemented, in the following we consider system sizes which are not in this asymptotic limit, and therefore we still allow for a full contraction of the local energy contributions with moderate computational resources.
However, though this thresholding is therefore not used, this reduction of the asymptotic scaling can nevertheless become a helpful tool for pushing the approach to larger systems which would otherwise be inaccessible.}

\subsection{Practical considerations of the GPS}
Having set up the computational basis of the problem, we describe the quantum state as a GPS associating amplitudes with basis states according to Eq.~\eqref{eq:qGPS}.
We optimize the parametrized ansatz by minimization of the stochastically approximated energy expectation value with the standard Stochastic Reconfiguration method~\cite{PhysRevB.64.024512, becca_sorella_2017} until convergence of the variational energy is observed.
For all the numerical tests discussed in the following, we used a moderate number of $\approx 10,000$ samples for the approximation of expectation values.
Our implementation is based on the {\tt NetKet} software package~\cite{Carleo_2019, Vicentini_2022}, with additional functionality, including the GPS model definition and the implementation of the \emph{ab initio} Hamiltonian, publicly available via the {\tt GPSKet} plugin library (with scripts to generate results in this paper included in the repository).
To set up the molecular orbitals and obtain the one- and two-electron integrals we utilized the {\tt PySCF} package~\cite{https://doi.org/10.1002/wcms.1340, doi:10.1063/5.0006074}.

A central element of the state approximation is the evaluation of the local energy for each sampled configuration, as defined in Eq.~\eqref{eq:localE}.
This requires the evaluation of the Hamiltonian matrix elements, ${\hat H}_{x, x'}$, as well as the amplitudes, $\Psi(x')$, for all configurations connected via a non-zero Hamiltonian matrix element to $\Psi(x)$.
As is presented in more detail in App.~\ref{sec:practical_evaluation}, the evaluation of the contributing terms can be performed efficiently for the GPS by considering local updates to pre-computed quantities.
This allows us to evaluate the amplitudes of a GPS ansatz for each connected configuration, $\Psi(x')$, via local updates, with a complexity of $\mathcal{O}[M]$, therefore independent of the system size.
The evaluation of the Hamiltonian matrix elements involves a computation of a parity prefactor ensuring the fermionic commutation relations depending on a chosen normal ordering of the orbitals~\cite{altlandCondensedMatterField2010a}.
The computation of parity prefactors is equivalent to the evaluation of Jordan-Wigner type mappings from fermions to spin/qubit degrees of freedom~\cite{doi:10.1021/acs.jctc.8b00450}, which we can also evaluate in constant time for each term in the Hamiltonian with appropriate setup.
Overall, each local energy evaluation therefore scales as $\mathcal{O}[M \times N^2 \times (L-N)^2]$, not taking into account any pruning of (approximately) vanishing terms.

The canonical basis directly builds upon mean-field simulations, making it trivial to recover the HF level of accuracy with a wavefunction for which all but one amplitude vanishes --- a distribution easily represented as a GPS.
In the local basis however, it is not immediately obvious how mean-field properties can be recovered with this ansatz, and we often found it difficult to reliably reach the uncorrelated approximation in our simulations with a GPS.
Furthermore, there are questions raised regarding the ordering of the fermionic degrees of freedom.

A specific ordering of the orbitals is required to define a normal order for the evaluation of parity prefactors in the Hamiltonian~\cite{altlandCondensedMatterField2010a}.
While we can define a natural choice to order the orbitals in a canonical basis via the single-particle energy level, ordering the orbitals becomes ambiguous and ill-defined for all but one-dimensional systems represented in a local basis.
However, we show in App.~\ref{sec:fermionic_ordering} that the effect of \emph{all} possible orbital reorderings on the sign structure can be efficiently captured in the span of the GPS model by increasing the support dimension of the model polynomially in system size (quadratically). Put another way, any GPS model with support dimension $M$, is able to have all possible observables reproduced under any fermionic orbital reordering, by a model with support dimension $M+\mathcal{O}[L^2]$. This is a manifestation of the non-local correlation in the amplitudes (in this case their sign) that the GPS model can describe, allowing this changing sign structure from fermionic orbital reordering to be expressed in a polynomially compact fashion.
However, the explicit construction suggests that, in general, we require a support dimension scaling quadratically with the number of orbitals to be able to span a state which is invariant to the choice of the ordering.
This is still a relatively high scaling, which would reduce the efficiency of the method. 

As a practical alternative, we can instead augment the GPS with an explicitly anti-symmetric reference state, such as a single Slater determinant (SD), allowing the effect of orbital reorderings to be entirely subsumed within this reference, and avoiding the ambiguities of orbital ordering without requiring an explicit scaling of the support dimension with system size. This can replicate the success of such constructions for fermionic lattice models~\cite{Nomura2017, Glielmo_2020, https://doi.org/10.48550/arxiv.2210.05871}, and furthermore allows us to incorporate the properties of the uncorrelated physics without an increase in the support dimension, or significant impact on the overall computational cost~\cite{doi:10.1063/1.4829835}. This also ensures that we can rigorously describe the mean-field character of the state, using the GPS in a similar spirit to the Jastrow factor in standard Slater-Jastrow ansatzes~\cite{becca_sorella_2017}.
However, by simultaneously optimizing the reference state and GPS, this construction does not limit the ability to (theoretically) approach exactness of the description by increasing the GPS support dimension~\cite{morenoFermionicWaveFunctions2021}.


\begin{figure}
    \includegraphics[width=\columnwidth]{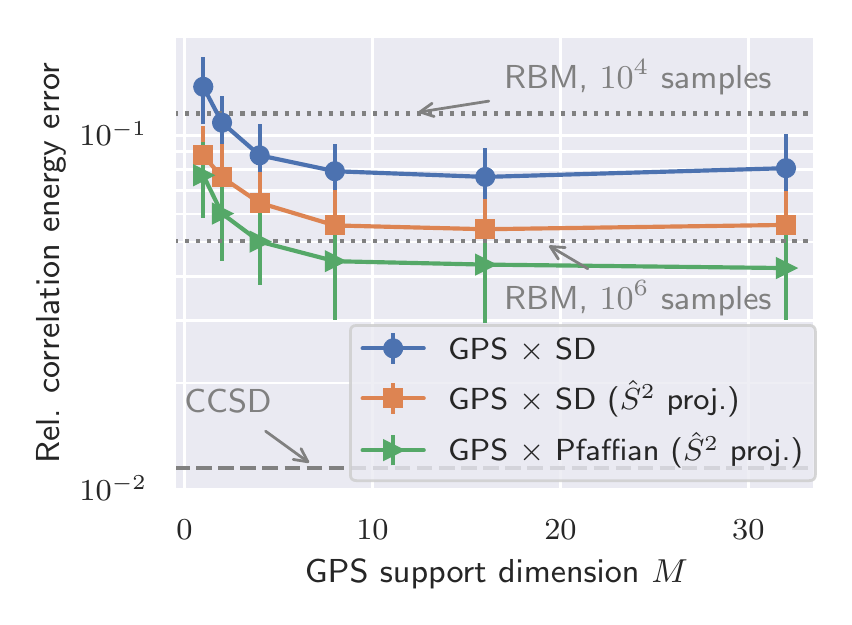}
    \caption{Relative error in the correlation energy \toadd{(w.r.t. the exact ground state energy)} obtained for a water molecule (6-31G basis, geometry as specified in Ref.~\onlinecite{Choo_2020}), in relation to the support dimension of the GPS which is augmented by different reference states. Co-optimized reference ansatzes include an unrestricted SD (blue circles), a spin-projected unrestricted SD (orange squares), and a spin-projected anti-parallel Pfaffian ansatz (green triangles). \toadd{The error bars were computed from the standard deviation of the variational energy over the last $50$ optimization steps, rescaled by the exact correlation energy.} Figure includes reference values achieved with an RBM, optimized with $10^4$ and $10^6$ samples, from Ref.~\cite{Choo_2020}, as well as a CCSD value.}
    \label{fig:H2O}
\end{figure}

As a simple example, we replicate the description of a water molecule in a 6-31G basis set, as discussed in Ref.~\cite{Choo_2020}.
That work discusses the restricted Boltzmann machine (RBM) neural network architecture\toadd{, comprising a single hidden layer of neurons,} as ansatz for the state.
\toadd{The number of hidden nodes in the network serves as the model's main hyperparameter equivalent to the support dimension in the GPS, controlling its flexibility and computational cost, which is comparable between the two models.}
It was shown that the achieved accuracy strongly depends on the number of Monte Carlo samples, which was attributed to a particularly peaked sampling distribution of the target, indicating the use of a canonical basis.
While it is one of the larger systems discussed in the {\em ab initio} study with RBM, the system still allows for a treatment with full configuration interaction techniques providing an exact baseline reference.
Fig.~\ref{fig:H2O} shows results achieved with a variationally optimized GPS, augmented by a reference state optimized alongside the GPS, as a function of the support dimension.
We present results for three different reference states: a single unrestricted SD, an $\hat{S}^2$-projected unrestricted SD, and a $\hat{S}^2$-projected anti-parallel Pfaffian state~\cite{PhysRevLett.96.130201, MISAWA2019447}.
While we were not able to achieve similar accuracies with a sole GPS, with the augmentation, our results mostly improve upon the accuracy reported for the RBM with only $\approx 10^4$ samples.

A single SD without explicit spin projection, together with the most simple GPS with a support dimension $M=1$, gives a corresponding relative correlation energy error of $\approx 14 \%$.
The error can be reduced further by either increasing the support dimension, thus adding further variational flexibility in the GPS part, or by allowing more flexibility in the reference state.
At $M=1$, the spin projected SD decreases the relative energy error by approximately $5 \%$, which is then further decreased by roughly another percent through utilization of the spin-projected Pfaffian.
Although, we see an improvement of the description for small support dimensions, the energetic values saturate for support dimensions larger than $M=8$.
With a full $\hat{S}^2$ projected Pfaffian reference state, the relative correlation energy error converges to a value of $\approx 4.2 \%$, marginally improving upon the \toadd{overall} best \toadd{reported} RBM result of Ref.~\cite{Choo_2020}, which was obtained with \toadd{the maximum considered number of} $10^6$ samples\toadd{, with an RBM hidden unit density of one hidden neuron per spin-orbital}.
Nonetheless, the results do not match the accuracy obtained from coupled cluster calculations based on single and double excitations (CCSD).
The lack of further accuracy improvements with the GPS of larger support dimensions suggests that additional optimization difficulties limit the manifestation of the systematic improvability suggested by the ansatz construction.
\toadd{We have confirmed that this limitation is, in fact, not caused by shortcomings in the stochastic approximation of expectation values, but also persists similarly if expectation values are evaluated from a contraction over the full computational basis without stochastic noise.
The lack of systematic improvability can therefore either be attributed to fundamental limitations of the model in the considered limit or to difficulties with faithfully finding the optimal model parameters, often identified as a notoriously hard challenge for machine learning inspired ansatzes which can suffer from restrictive parameter landscapes or generalization difficulties~\cite{Szab__2020, Bukov_2021, frank2021learning}.
Although more than $90\%$ of the correlation energy is captured in our results, further practical improvements to the algorithm are required to reach arbitrary accuracies to match the high level of accuracy obtained from coupled cluster calculations for this system.
However,} whereas coupled cluster approaches are particularly successful for systems exhibiting relatively weak degrees of electronic correlation, the GPS model does not particularly target this limit, and we expect a more general applicability of the model.

\section{Towards extended hydrogen materials}
\label{sec:results}
To test the assertion of applicability in more strongly correlated \emph{ab initio} systems, we benchmark the methodology for simple arrays of hydrogen atoms, described in a minimal basis set representation, already giving rise to rich quantum phenomena of condensed matter systems driven by strong electronic correlation.
Conceptually, such hydrogen materials share a high degree of similarity to Fermi-Hubbard models, as well as the ability to change physical correlation regimes via changes in bond lengths. However, these hydrogen models are extended to general quartic electron interactions, as well as single-particle Hamiltonian terms, which range across the full system.
They have therefore become a common testing-ground for electronic structure methods~\cite{PhysRevX.7.031059, Stella_2011, doi:10.1063/1.3459059, doi:10.1063/1.3237029, Hachmann_2006}, for which we benchmark the ability of the GPS ansatz to capture the strong electronic-correlation emerging in these systems as the inter-atomic spacing is increased, whilst ensuring that the interactions remain more faithful to the true Coulombic one.

\begin{figure}
    \includegraphics[width=\columnwidth]{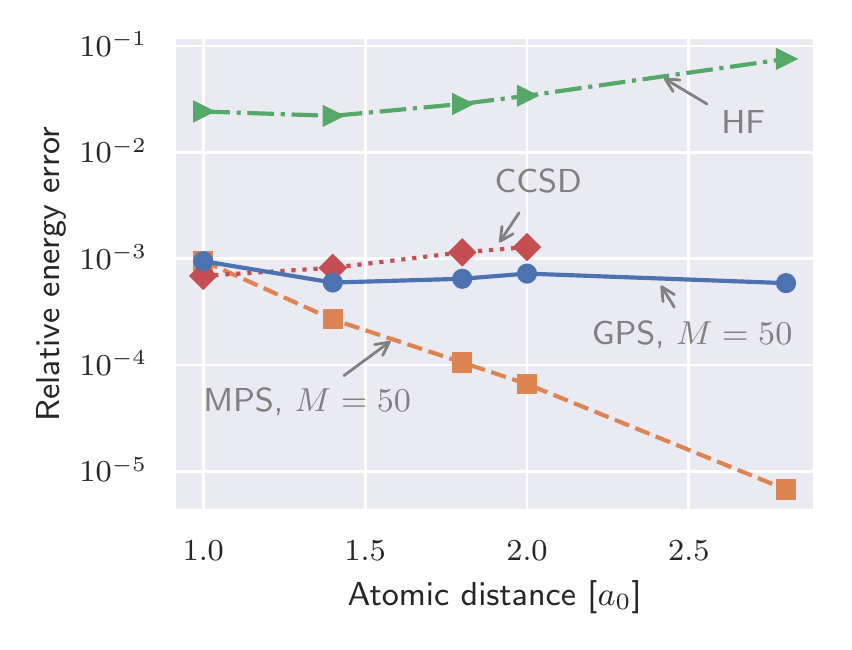}
    \caption{Relative energy errors for linear hydrogen chains of $50$ atoms at different atomic separations as obtained by different approaches. Data points include results from HF (green triangles), the GPS description as outlined in the main text (blue circles), CCSD (red diamonds), and DMRG calculations with fixed bond dimension of $M=50$ (orange squares). Errors are evaluated by comparison to DMRG results converged with the MPS bond dimension. DMRG, HF, and CCSD results taken from Ref.~\onlinecite{Hachmann_2006}.}
    \label{fig:H50}
\end{figure}

In Fig.~\ref{fig:H50} we describe obtained accuracies for a one-dimensional hydrogen chain comprising $50$ atoms at different inter-atomic separations.
The quasi-one-dimensionality of the system limits its entanglement and makes it possible for DMRG to provide essentially exact descriptions for these systems in a local basis representation, and we compare our results to DMRG results with a converged MPS bond dimension from Ref.~\onlinecite{Hachmann_2006}.
The quasi-one-dimensional nature also avoids complicated nodal structures in the described wavefunction solely emerging due to fermionic ordering ambiguities, and we were able to achieve competitive accuracies solely with a GPS of practically manageable support dimension, which we chose as $M=L=50$, not requiring the inclusion of a reference state to capture mean-field characteristics and avoid orbital ordering ambiguities.
In addition to the GPS results, the figure also includes results obtained with HF, CCSD results (where calculations could be converged), and DMRG results with fixed MPS bond dimension of $M=50$, all taken from Ref.~\cite{Hachmann_2006}.

The electronic correlation contributes significantly to the physical characteristics of the system.
This manifests in an inability to reach reasonable accuracy with HF methods, giving relative energy errors greater than $1 \%$ for all considered separations, which goes up to $\approx 7.6 \%$ for an atomic separation of $2.8 \, a_0$ as the correlations become more significant.
The optimization of the GPS model within the VMC framework, on the other hand, consistently reaches an error of slightly less than $0.1 \%$ for all geometries.
This level of accuracy is mostly in agreement with that achieved from converged CCSD calculations at equilibrium and weaker correlation regimes.
Importantly, however, the GPS also reaches this accuracy for the largest considered separation where the CCSD calculation could not be converged, indicating a good consistency of the GPS across different physical regimes.
Interestingly, the relative energy error of the GPS matches that of an MPS with bond dimension equal to the GPS support dimension at $1 \, a_0$, despite the significantly smaller number of variational parameters in the GPS.
Nevertheless, the relative energy error from the MPS decreases as the geometry becomes more stretched, indicating a decay of entanglement rank between the orbitals.
While the GPS does not follow this accuracy improvement, being able to reach a consistent level of accuracy across different correlation regimes is a good indication of the model's general ability to compress the target state efficiently.

\begin{figure}
    \includegraphics[width=\columnwidth]{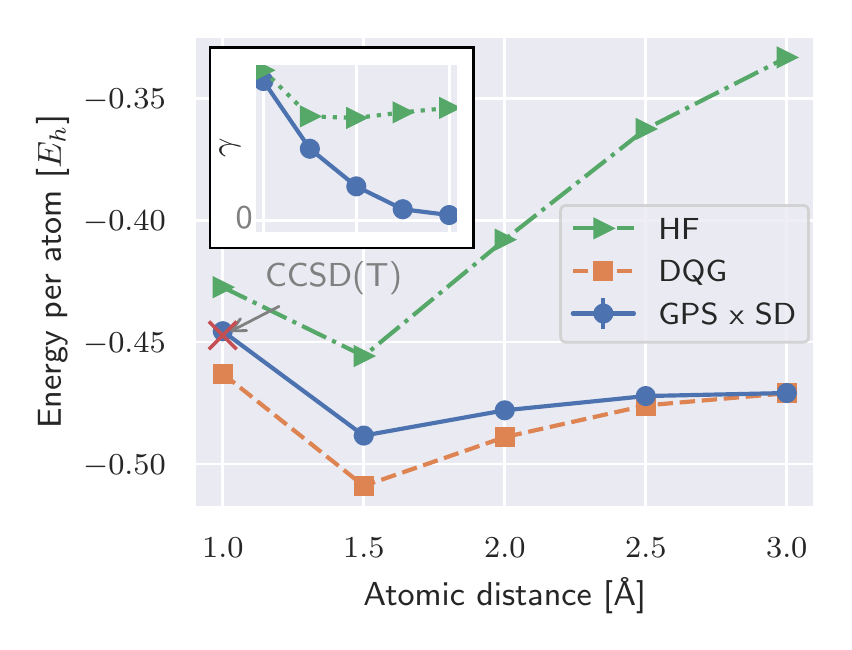}
    \caption{Results for a cubic system of $4\times 4 \times 4$ hydrogen atoms (STO-6G atomic orbital basis) at different inter-atomic distances. Main panel shows energies per atom obtained with GPS ($M=96$) multiplied by an unrestricted SD (blue circles), as well as HF results (green triangles), a single CCSD(T) value (red cross), and DQG results taken from Ref.~\onlinecite{doi:10.1063/1.3459059} (orange squares). Inset shows the electronic mobility coefficient $\gamma$ as defined in Eq.~\eqref{eq:gamma} from the VMC optimized state and the HF wavefunction.}
    \label{fig:H_cube}
\end{figure}

While the MPS description explicitly builds on a one-dimensional structure to define an exactly contractible representation, the stochastically optimized GPS is not explicitly tailored for one-dimensional systems, also enabling efficient descriptions for higher-dimensional systems~\cite{Glielmo_2020, Rath_2022}.
By augmenting the pure GPS with an explicitly anti-symmetric reference state for fermionic systems, the description becomes entirely independent to the imposed fermionic orbital ordering. Furthermore, it allows us to capture mean-field effects efficiently, while allowing for systematic improvements of the variational flexibility.
To highlight the ability of the augmented model to simulate systems for which no numerically exact methods are available, we study a cubic array comprising $4 \times 4 \times 4$ hydrogen atoms, and simultaneously stretch all bonds symmetrically, breaking all bonds simultaneously.
This constitutes a system size beyond what has previously been tackled with comparable NQS studies, and a careful implementation of the methodology allows us to present new state-of-the-art benchmarks for this challenging system with long-range interactions and tuneable correlation strength. While it is still somewhat contrived (especially due to the limited basis size), it is an important first step towards realistic periodic and extended systems with this methodology.
We report results achieved with a GPS of fixed support dimension, $M=96$, optimized together with a SD of fixed magnetization, in Fig.~\ref{fig:H_cube}.
\toadd{Based on our experience, we expect that such a support dimension of the order of the number of orbitals, allows us to reach a high accuracy level beyond which diminishing returns are found with additional increase in the support dimension.}

In the main panel of the figure, we show the energy per atom as the atomic separation is varied.
The optimized GPS ansatz predicts a similar equilibrium geometry as obtained from the HF baseline, also indicated in the figure, giving an energy minimum for an atomic spacing of about $1.5 \,${\AA}.
Crucially, however, we are able to achieve results significantly improving upon the HF level of accuracy, with an increasingly large discrepancy between the augmented GPS and the HF energies as the atomic separations get larger, resulting in a significant reduction of the harmonic frequency of the symmetric vibrations about equilibrium due to the correlation.
With the ability to model local correlation properties with the GPS, we observe the expected asymptotic convergence of the energy as the cluster is dissociated, which cannot be captured based on mean-field considerations lacking the required charge fluctuations.

To compare the accuracy of the obtained energy values, the figure also includes the energy values obtained from a variational two-body reduced density matrix approach with approximate $N$-representability enforced via the `DQG' conditions, as discussed in Ref.~\cite{doi:10.1063/1.3459059}.
While the VMC framework always produces an upper bound to the exact ground state energy, the DQG energies represent a lower bound to this value.
Both methods give good agreement in the limit of large atomic separation, predicting an energy per atom of $\approx - 0.471 \, E_h$ at $3 \,${\AA} spacing between the atoms, confirming this as an accurate approximation of the energy.
For less stretched geometries however, we obtain an increasing discrepancy between the two approaches.
While it was not possible to converge coupled cluster calculations for larger separations, at a distance of $1 \,${\AA}, we obtained an energetic comparison value from CCSD with perturbative inclusion of triple excitations (CCSD(T)) of $\approx -0.447 \, E_h$ per atom.
Although this is marginally smaller than the one suggested by our VMC calculation, its agreement with our VMC value is significantly better than with the DQG result, which increases our confidence that we obtained a highly accurate approximation with the GPS.

Going beyond the evaluation of the energy expectation values, we also confirm that the system undergoes a phase transition from a metal to a Mott-insulator as we increase the separation between hydrogen atoms.
In keeping with the analysis of Ref.~\onlinecite{doi:10.1063/1.3459059}, we characterize this transition by quantifying the instantaneous electronic mobility from the coherences of the one-body reduced density matrix.
More specifically, we evaluate the root mean square of its off-diagonal in a local (atomic) orbital basis, given by
\begin{equation}
    \label{eq:gamma}
    \gamma = \sqrt{\frac{\sum_{a \neq b = 1}^{2L} \sum_{i,j=1}^{2L} C_{a, i} \, C_{b, j} |\langle \Psi | \hat{c}^\dagger_i \hat{c}_j | \Psi \rangle |^2}{2L \times (2L -1)}},
\end{equation}
where the coefficients $C_{a, i}$ represent the change from the molecular basis with orbitals labeled by $i$ to the atomic basis labeled by index $a$, and the expectation values $\langle \Psi | \hat{c}^\dagger_i \hat{c}_j | \Psi \rangle$ are again evaluated via stochastic sampling.
We report the electronic mobility coefficient for the dissociation of the cubic hydrogen material from our simulations in the inset of Fig.~\ref{fig:H_cube}.
We observe a decay of the electronic mobility to zero, not captured on the mean-field level of accuracy.
Being able to predict this breakdown of instantaneous electron transfers induced through quantum many-body interactions underlines the applicability of the method to understand and describe quantum phenomena with realistic interactions driving technologically relevant material properties in which few, if any, reliable alternative approaches exist.

\section{Conclusions and outlook}
With the unparalleled progress of AI technology in the recent years, the exploitation of dualities and synergies with quantum many-body problems will likely provide an increasingly important research area in the near future. 
This work provides further exemplification of how powerful ML frameworks can provide tools to extract quantum physical properties from fundamental principles.
We have highlighted the general applicability of modern ML-inspired functional forms in a second quantized VMC framework for chemical predictions from {\em ab initio} principles.
In particular, we considered the choice of a local basis for alleviation of practical difficulties in the VMC optimization driven by stochastic sampling from the computational basis.
While mean-field characteristics are implicitly incorporated in a canonical molecular orbital basis, if required, we incorporate these by inclusion of appropriate reference states within a local basis representation.
We have exemplified the methodology utilizing the recently developed GPS ansatz, corresponding to a particularly simple functional form inspired by kernel models and Bayesian regression, controlled by a single hyperparameter to describe the expressiveness in the constructed chosen computational basis.
Providing a new benchmark level of accuracy, and going beyond system sizes previously studied with comparable approaches, we describe a metal-to-insulator transition driven by quantum correlations in a three-dimensional hydrogen material comprising $64$ atoms.
The description in a local basis, for which a natural sparsity of the Hamiltonian emerges, also provides natural extensions of the approach to enable larger scale simulations.

Based on a variety of recent benchmarks for prototypical lattice models, we expect the presented results to be largely independent of the chosen ML-inspired functional form to define the ansatz.
The discussed framework is, amongst others, equivalently applicable to similarly motivated ansatzes constructed from neural network representations\toadd{, such as the RBM}, which we anticipate to reach comparable results.
While the definite confirmation of this assumption requires further benchmarks, there is increasing evidence that results achievable with highly expressive variational functional forms in a VMC context are often limited by shortcomings of the optimization procedures rather than the model's expressivity~\cite{Westerhout_2020, Bukov_2021}.
Indeed, we were practically not able to observe a general improvability of the model to arbitrary target accuracies for fermionic systems.
While the theoretical expressivity of the GPS can be improved by increasing its support dimension, even for a relatively simple testing system we observed a saturation to an accuracy limit beyond which no further improvements materialized.
We therefore particularly consider further developments of techniques to overcome such limitations to be of major importance in order to extend the abilities of ML-inspired formalisms to study chemical properties from {\em ab initio} simulations in a second quantized framework.
This might be achieved through modifications to the optimization strategies~\toremove{\cite{https://doi.org/10.48550/arxiv.1811.12423, https://doi.org/10.48550/arxiv.2211.07749}}\toadd{\cite{https://doi.org/10.48550/arxiv.1811.12423, Roth2022, lovato2022hiddennucleons, chen2023efficient}} or by considering different paradigms to construct the electronic state, e.g., by following the construction of fully-flexible, explicitly anti-symmetrized representations~\cite{Luo2019, morenoFermionicWaveFunctions2021} as these are commonly applied with great success in the real space picture~\cite{PhysRevResearch.2.033429,Hermann_2020,spencerBetterFasterFermionic2020,https://doi.org/10.48550/arxiv.2205.09438,hermannAbinitioQuantumChemistry2022, liInitioCalculationReal2022, vonglehnSelfAttentionAnsatzAbinitio2022, Han2018, https://doi.org/10.48550/arxiv.2302.04168, https://doi.org/10.48550/arxiv.2110.05064, https://doi.org/10.48550/arxiv.2205.14962}.

\section*{Code Availability}

The code used in this work can be found at \url{https://github.com/BoothGroup/GPSKet}, with the input and scripts to generate all results available at \url{https://github.com/BoothGroup/GPSKet/tree/master/scripts/GPS_for_ab_initio}

\begin{acknowledgments}
The authors gratefully acknowledge support from the Air Force Office of Scientific Research under award number FA8655-22-1-7011, as well as the European Union's Horizon 2020 research and innovation programme under grant agreement No. 759063. We are grateful to the UK Materials and Molecular Modelling Hub for computational resources, which is partially funded by EPSRC (EP/P020194/1 and EP/T022213/1).
Furthermore, we acknowledge the use of the high performance computing environment CREATE at King's College London~\footnote{King's College London. (2022). King's Computational Research, Engineering and Technology Environment (CREATE). Retrieved January 13, 2023, from \url{https://doi.org/10.18742/rnvf-m076}}.
\end{acknowledgments}

\newpage
\appendix
\section{Fast evaluation of local energy terms for GPS}
\label{sec:practical_evaluation}

In order to evaluate the local energy for a configuration $| x \rangle$ from the Hamiltonian in Eq.~\eqref{eq:el_structure_ham}, amplitudes arising from the model due to one-(two-)electron operators of the form
\begin{equation}
    E_{i,j,(k,l)} = \frac {\langle x | \hat{c}^\dagger_i (\hat{c}^\dagger_k \hat{c}_l) \hat{c}_j | \Psi \rangle}{\langle x | \Psi \rangle}
\end{equation}
need to be evaluated.
The indices $i,j,k,l$ label different spin-orbitals, and the additional operators $\hat{c}^\dagger_k \hat{c}_l$ are only present for two-electron terms.

For a model associating amplitudes to basis states from the Fock basis, the evaluation of the terms generally involve two steps.
Firstly, the connected Fock state, $| x' \rangle = P_{x,x'} \, \hat{c}^\dagger_j (\hat{c}^\dagger_l \hat{c}_k) \hat{c}_i \, | x \rangle$ needs to be identified.
This involves the evaluation of a parity prefactor $P_{x,x'}$, emerging through the fermionic commutation relations.
The value of $P_{x,x'}$ is zero if the final electronic occupancy in every degree of freedom does not agree with that of $| x \rangle$.
Otherwise, it is plus or minus one, depending on the number of electrons which are passed by the (double) electron move according to the chosen ordering of orbitals~\cite{altlandCondensedMatterField2010a}.
The identification of the connected configuration and the evaluation of the parity prefactor are equivalent to the evaluation of Pauli operator strings in a Jordan-Wigner transformation from fermionic degrees of freedom to spin/qubit systems~\cite{doi:10.1021/acs.jctc.8b00450}.
With the identification of a connected Fock state, the local energy contribution is given as
\begin{equation}
    E_{i,j,(k,l)} = P_{x,x'} \frac{\langle x' | \Psi \rangle}{\langle x | \Psi \rangle}
    \label{eq:local_en_contribution}
\end{equation}
and therefore also requires the evaluation of the amplitude ratio $\frac{\langle x' | \Psi \rangle}{\langle x | \Psi \rangle}$ defined from the ansatz.

Due to the large number of terms which contribute to the local energy, it is often helpful to consider more efficient local updates to evaluate each term.
This commonly involves appropriate setup for the sample configuration $| x \rangle$ to enable a more efficient calculation for the connected configurations.
By storing the cumulative electron counts in the different orbitals in the setup, the parity prefactor, directly obtained through counting the number of passed electrons, can be evaluated in constant time for each connected configuration.

Similar to the fast update of Slater-Jastrow models~\cite{doi:10.1063/1.4829835}, we can also utilize update equations for the GPS ansatzes considered in this work.
These exploit the fact that each connected configuration $| x' \rangle$ gives an electronic occupancy which differs by at most the occupancy of four orbitals compared to the sampled configuration $| x \rangle$.
To evaluate the corresponding amplitude update for the GPS as specified in Eq.~\eqref{eq:qGPS} of the main text, its amplitude, evaluated for configuration $| x \rangle$, can be represented as
\begin{equation}
    \Psi(x) = \exp \left(\sum_{\alpha=1}^M \varphi_{\alpha}(x) \right).
\end{equation}
Here, $\varphi_{\alpha}(x)$ denote unnormalized product state amplitudes defined via the variational parameters of the GPS, $\epsilon$, as a product over the spatial orbitals according to $\varphi_{\alpha}(x) = \prod_{i=1}^L \epsilon_{\alpha, i, x_i}$.
Using pre-computed product state amplitudes, $\varphi_{\alpha}(x)$, the amplitude for the connected configuration evaluates to
\begin{equation}
    \Psi(x') = \exp \left(\sum_{\alpha=1}^M \varphi_{\alpha}(x) \times \prod_{\bar{i} \in \{i,j,(k,l)\}} \left (\frac{\epsilon_{\alpha, \bar{i}, x'_{\bar{i}}}}{\epsilon_{\alpha, \bar{i}, x_{\bar{i}}}} \right) \right),
\end{equation}
where the set $\{i,j,(k,l)\}$ contains (at most) four indices, labelling the orbitals with changed occupancy.
Using the pre-computed amplitude of the central configuration, $\Psi(x)$, as well as the $M$ product state amplitudes, $\varphi_{\alpha}(x)$, each term in the local energy can therefore be evaluated in $\mathcal{O}[M]$ time.
This scaling, independent of the number of orbitals, is generally unaffected by the inclusion of a mean-field type reference state, for which similar low-rank updates can be performed~\cite{doi:10.1063/1.4829835}.

\section{Orbital reordering invariance of GPS}
\label{sec:fermionic_ordering}
While the span of representable amplitudes (including sign) is independent of the imposed ordering of the orbitals for a GPS, the quality of results (in the absence of a reference state) will still depend on this chosen ordering for fermionic systems.
This is due to the fact that the sign structure of the modeled target amplitudes in the Fock basis changes under a change of the ordering, due to the different parity prefactors $P_{x, x'}$ in Eq.~\eqref{eq:local_en_contribution}.
We show in this appendix that we can construct a GPS with support dimension scaling at most quadratically with the system size which is able to express {\em all} changes induced in the sign structure of a fermionic state due to a different normal ordering.

For two different choices to define the normal ordering, basis states from the original computational basis can be related to basis states from a basis with changed orbital ordering via a sign transformation.
Assuming a particular choice of molecular orbitals and letting $|x \rangle$ be the states from the associated computational basis, we express the basis states as
\begin{equation}
    |x \rangle = c^\dagger_{r(0)} c^\dagger_{r(1)} \ldots c^\dagger_{r(N)} | 0 \rangle.
    \label{eq:creation_string}
\end{equation}
Here, $r(i)$ labels the spin-orbital which is occupied by electron with index $i$, and we choose the electron labels such that the indices satisfy $r(0) < r(1) < \ldots < r(N)$ (we make the choice of sorting the electrons such that all the spin-up electrons are followed by all the spin-down electrons).
While there might be a natural choice of this ordering for some systems and molecular orbital choices (e.g., by energy level in the canonical basis or by position in space for one-dimensional systems in a local basis), ambiguities emerge in other cases, e.g. for larger basis sets, or local degrees of freedom in more than one dimension.

To compare the sign structure emerging through a reordering of the orbitals, we consider a relabeling of the orbitals to give a different set of Fock basis states,
\begin{equation}
    |\tilde{x} \rangle = c^\dagger_{\tilde{r}(0)} c^\dagger_{\tilde{r}(1)} \ldots c^\dagger_{\tilde{r}(N)} | 0 \rangle.
\end{equation}
The electron labels are again chosen to fulfill a normal ordering, $\tilde{r}(0) < \tilde{r}(1) < \ldots < \tilde{r}(N)$, however, now defined w.r.t a different linear order.
This is specified by a permutation $\mathcal{P}$ defining a one-to-one mapping $\mathcal{P}(r) = \tilde{r}$ from an index $r$ to a new index $\tilde{r}$ labeling the same physical orbital.
Each new basis state can be associated with one from the original basis multiplied by a configuration-dependent sign according to
\begin{equation}
    |x \rangle = (-1)^{\mathcal{N}_{x,\tilde{x}}} |\tilde{x} \rangle.
\end{equation}
The sign relating the two basis states is simply the parity of the permutation exchanging the creation operators in the definition of the basis states from one order into the other.
It can be expressed as $(-1)^{\mathcal{N}_{x,\tilde{x}}}$, where $\mathcal{N}_{x,\tilde{x}}$ is the number of pairwise electron exchanges required to reassign the electron labels so that the list of orbital indices $(\tilde{r}(0), \tilde{r}(1), \ldots, \tilde{r}(N))$ satisfies the order $\mathcal{P}^{(-1)}(\tilde{r}(0)) < \mathcal{P}^{(-1)}(\tilde{r}(1)) < \ldots < \mathcal{P}^{(-1)}(\tilde{r}(N))$.

The additional sign structure, $(-1)^{\mathcal{N}_{x,\tilde{x}}}$, in the wavefunction amplitudes, solely emerging through a reordering of the orbitals, can be represented as a GPS with support dimension $M$ scaling at most quadratically in the number of molecular orbitals.
To show this, we decompose the number of pairwise exchanges, $\mathcal{N}_{x,\tilde{x}}$, as a sum over all pairwise next-element exchanges in the permutation $\mathcal{P}$, according to
\begin{equation}
    \mathcal{N}_{x,\tilde{x}} = \sum_{\sigma \in \{\uparrow, \downarrow\}} \sum_{\{(a, b)\}} n_{a, \sigma}(x) n_{b, \sigma}(x).
    \label{eq:number_exchanges}
\end{equation}
Here, the $(a,b)$ sum runs over all index pairs which need to be exchanged so that the original list of orbital indices $(1, \ldots, L)$ is iteratively brought into the order $(\mathcal{P}(1), \ldots \mathcal{P}(L))$ by only exchanging indices which are directly adjacent.
The occupation number $n_{i, \sigma}(x)$ gives the number of electrons occupying spin channel $\sigma$ of spatial orbital $i$ in the many-electron configuration $|x\rangle$, thus either evaluating to one or zero.
If (and only if) the configuration $| x \rangle$ has an electron with the same spin in both of the orbitals labeled by $a$ and $b$ then two creation operators in the construction of Eq.~\eqref{eq:creation_string} are exchanged, resulting in an additional $(-1)$ prefactor from the commutation relations.
Iteratively applying the pairwise exchanges of adjacent creation operators according to the permutation of orbitals $\mathcal{P}$ then yields the desired representation together with the induced sign transformation.

The representation of the sign transformation as a GPS follows directly from the representation of $\mathcal{N}_{x,\tilde{x}}$ according to Eq.~\eqref{eq:number_exchanges}.
Specifically, we can represent the sign structure $(-1)^{\mathcal{N}_{x,\tilde{x}}}$ as a GPS by associating each support point index, $\alpha$, with a term from Eq.~\ref{eq:number_exchanges}, i.e., a unique pair of orbital indices from the set $\{(a,b)\}$ and associated spin value $\sigma \in \{\uparrow, \downarrow\}$.
Based on the definition of the GPS amplitudes according to Eq.~\eqref{eq:qGPS}, the representation is, e.g., obtained with the parameter choice
\begin{align}
    \epsilon_{\alpha, a, \sigma} = \epsilon_{\alpha, a, \uparrow \downarrow} &= i \pi \nonumber \\
    \epsilon_{\alpha, b, \sigma}  = \epsilon_{\alpha, b, \uparrow \downarrow} &= 1 \nonumber \\
    \epsilon_{\alpha, a, \bar{\sigma}} = \epsilon_{\alpha, b, \bar{\sigma}} = \epsilon_{\alpha, a, \cdot} = \epsilon_{\alpha, b, \cdot} &= 0 \nonumber \\
    \epsilon_{\alpha, i \not \in \{(a,b)\} , l} &= 1 \nonumber,
\end{align}
where $\bar{\sigma}$ denotes the inversion of spin $\sigma$ and index $l$ runs over all possible local occupancies, $l \in \{\uparrow, \downarrow, \uparrow \downarrow, \cdot\}$.

With this construction, the GPS representation of the sign structure relating two different orbital orderings therefore requires a support dimension of $M = 2 |\{(a,b)\}|$, where $|\{(a,b)\}|$ corresponds to the number of pairwise index exchanges in the permutation of indices specified by $\mathcal{P}$.
Any such permutation comprises at most $\mathcal{O}(L^2)$ pairwise exchanges, consequently limiting the required support dimension (and therefore the total number of variational parameters) of the constructed GPS to scale at most quadratically with the number of orbitals.
This also means that we can always define a GPS with a support dimension $M$ increased by at most $\mathcal{O}(L^2)$ so that it exactly spans any other GPS with support dimension $M$ with a changed ordering of the molecular orbitals in the definition of the basis states.
Though the span of a GPS with given support dimension is therefore not fully  invariant under changes to the orbital ordering, it always contains a subset of states which can be represented independent of the orbital ordering if the support dimension is scaled quadratically with the system size.
Nonetheless, an exact invariance of the state, independent of the support dimension, is also obtained through the inclusion of an explicitly anti-symmetrized reference state acting in a first-quantized picture, such as a single Slater determinant.


\bibliography{main}
\end{document}